\documentclass[prx, aps, twocolumn, floatfix, superscriptaddress, amsmath,  longbibliography]{revtex4-2}
\usepackage{graphicx}
\usepackage{graphics}
\usepackage{mathtools}
\usepackage{float}
\usepackage{amssymb}
\usepackage{bbold}

\usepackage{siunitx}

\DeclareSIUnit{\um}{\micro\meter}

\usepackage[dvipsnames]{xcolor}
\definecolor{darkblue}{rgb}{0.0, 0.0, 0.75}

\usepackage{color}
\usepackage[colorlinks=true,
            linkcolor=darkblue,
            urlcolor=darkblue,
            citecolor=darkblue]{hyperref}

	\usepackage[normalem]{ulem} 
	\definecolor{mgreen}{RGB}{1,123,0}

\def \br{{\bf r}}

\def \ms{\text{s}}
\def \mw{\text{w}}
\def \mb{\text{b}}
\def \mh{\text{h}}

\def \ms{\mathrm{s}}

\def \mHz{\mathrm{Hz}}

\def \mms{\mathrm{ms}}

\begin{document}
\title{Weak-link to tunneling crossover in an atomic Josephson junction}
\author{Vijay Pal Singh}
\affiliation{Quantum Research Centre, Technology Innovation Institute, Abu Dhabi, UAE}
\author{Erik Bernhart}
\affiliation{Department of Physics and Research Center OPTIMAS, RPTU University Kaiserslautern-Landau, 67663 Kaiserslautern, Germany}
\author{Marvin R\"ohrle}
\affiliation{Department of Physics and Research Center OPTIMAS, RPTU University Kaiserslautern-Landau, 67663 Kaiserslautern, Germany}
\author{Herwig Ott}
\affiliation{Department of Physics and Research Center OPTIMAS, RPTU University Kaiserslautern-Landau, 67663 Kaiserslautern, Germany}
\author{Ludwig Mathey}
\affiliation{Zentrum f\"ur Optische Quantentechnologien and Institut f\"ur  Quantenphysik, Universit\"at Hamburg, 22761 Hamburg, Germany}
\affiliation{The Hamburg Centre for Ultrafast Imaging, Luruper Chaussee 149, Hamburg 22761, Germany}
\author{Luigi Amico}
\affiliation{Quantum Research Centre, Technology Innovation Institute, Abu Dhabi, UAE}
\affiliation{Dipartimento di Fisica e Astronomia, Universit\`a di Catania, Via S. Sofia 64, 95123 Catania, Italy}
\affiliation{INFN-Sezione di Catania, Via S. Sofia 64, 95127 Catania, Italy}

\date{\today}
%

%
\begin{abstract}
We present a unified, quantitative description of transport across the crossover between hydrodynamic weak-link flow and tunneling-dominated Josephson dynamics in a three-dimensional quantum fluid. Using an atomic Josephson junction realized in a Bose–Einstein condensate, we continuously tune the barrier strength to access both regimes within a single, well-controlled system.
Measurements of the critical current and Josephson oscillations are in quantitative agreement with numerical simulations and analytical modeling, enabling a consistent inference of the microscopic mechanisms governing dissipation. In the weak-link regime, dissipative transport is consistent with vortex-ring–mediated phase slips, whereas in the tunneling regime it is consistent with rarefaction-pulse excitations.
The crossover is further reflected in a transition from a multi-harmonic to a predominantly single-harmonic current–phase relation, signaling the emergence of tunneling-dominated transport.
These results establish a general framework linking nonlinear excitations to coherent quantum transport across distinct dynamical regimes. More broadly, they provide insight into the microscopic origin of dissipation in driven quantum fluids, a problem that remains difficult to access in conventional solid-state systems.
\end{abstract}
\maketitle
%
%

\section{Introduction}

Josephson junctions provide a paradigmatic platform for studying coherent quantum transport \cite{Josephson1962}. Transport below the critical current is dissipationless and  exceeding this threshold leads to a resistive regime associated with the emergence of dissipation. 
Although coherent transport has been widely explored in both weak links and tunnel junctions \cite{likharev1979superconducting,sohn2013mesoscopic,golubov2004current}, accessing the crossover regime within a single solid-state platform remains essentially unfeasible.
Moreover, tracking the underlying microscopic dynamics — namely, identifying the excitations responsible for dissipation and understanding how their role evolves from weak-link transport to tunneling-dominated regimes — remains an open and largely unresolved problem \cite{Anderson1964, Grimes1966, Zimmerman1966, Hansma1971, Simmonds2001, Davis2002, Sukhatme2001, Hoskinson2006}.

Ultracold atomic gases provide a uniquely suitable platform to address this question. 
Atomic Josephson junctions enable precise control over geometry, interactions, and barrier strength, while allowing direct observation of the system's dynamics 
\cite{Cataliotti2001,Albiez2005, Levy2007, LeBlanc2011, Spagnolli2017,  Pigneur2018, Luick2020,Kwon2020, Pace2021, Rydow2025}, supported by extensive theoretical studies \cite{Smerzi1997,Raghavan1999,Giovanazzi2000, Meier2001, Kohler2003, Eckardt2005, Grond2011,Singh2020jj, Zaccanti2019,Xhani2020, Xhani2020NJP, Saha2021,  Zhu2021, Gabriel2023, Jahrling2025}. 
In contrast to solid-state implementations, the relevant timescales are sufficiently slow to 
permit real-time tracking of transport and dissipation processes, providing access to the underlying  microscopic dynamics \cite{SinghShapiro, DelPace2025,Bernhart2025,Singh2025}.

In this work, we experimentally and theoretically characterize the transition from hydrodynamic weak-link transport to the tunneling Josephson regime in a  Bose-Einstein condensate (BEC) of $^{87}$Rb atoms confined in a three-dimensional (3D) tube-like cylindrical geometry with a localized optical barrier. 
To this end, we  tune the barrier height, and follow the  crossover between these regimes  through measurements of the critical current and Josephson oscillations. 
Combining experiments with numerical simulations and analytical modeling, we infer the microscopic mechanisms governing dissipation across this crossover: 
In the weak-link regime, the dissipative transport is associated with vortex-ring excitations that 
drive phase slips and generate a finite chemical-potential difference across the junction; 
increasing the barrier height,  suppresses these hydrodynamics excitations and 
yields a tunneling-dominated Josephson regime in which dissipation occurs  consistently with the formation of rarefaction-pulse excitations. 
This crossover is further accompanied by a transition from a multi-harmonic to a predominantly single-harmonic current–phase relation, signaling the onset of tunneling-dominated transport.
This way, we formulate a unified and quantitative framework characterizing  dissipation across the weak-link–to–tunneling crossover and we offer key insights into the microscopic processes governing transport in 3D atomic Josephson junctions.

\begin{figure}[t]
\includegraphics[width=1.0\linewidth]{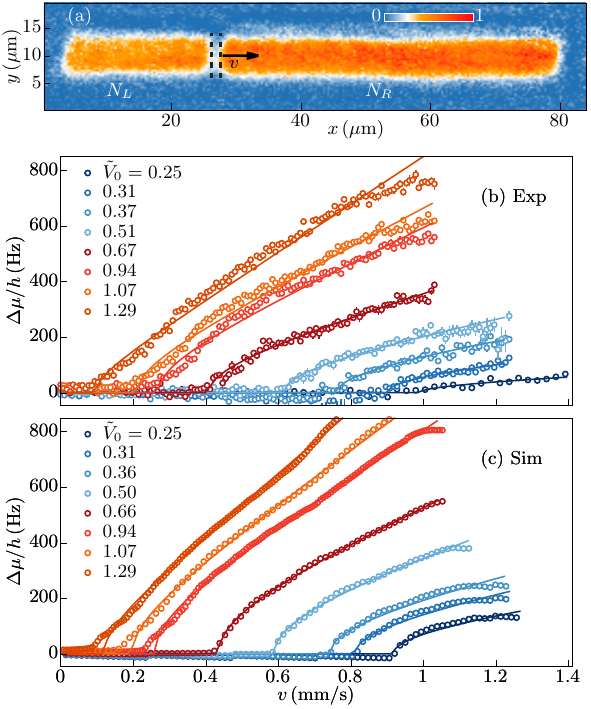}
\caption{Atomic Josephson junction in a trapped 3D Bose gas.  
(a) Absorption image of the atomic cloud together with an optical barrier (indicated by dashed lines) that moves at a constant velocity $v$ 
along the $x$ direction for $33\, \mms$. 
The barrier width is $w \approx \SI{1.1}{\um}$, and the barrier height $V_0$ is tunable. 
The barrier effectively separates the cloud into two subclouds with atom numbers $N_L$ and $N_R$. 
(b) Measured  velocity-chemical potential $(v-\Delta \mu)$ characteristics at different barrier heights $\tilde{V}_0 \equiv V_0/\mu$, 
where $\mu$ is the trap-averaged chemical potential.  
$\Delta \mu$ is determined from the atom number imbalance $z= (N_R-N_L)/(N_L+N_R)$; see text. 
The $y$ error bar denotes the statistical error.  
(c) Simulation of the $v-\Delta \mu$ characteristics using experimental parameters at temperature $T=23\, \mathrm{nK}$.   
The results are fitted with $\Delta \mu = \sqrt{v^2 -v_c^2}/G$ (solid lines), 
where the critical velocity $v_c$ and the conductance $G$  are the fit parameters. 
}
\label{Fig:system}
\end{figure}

\section{Experimental setup}
\begin{figure}[t]
\includegraphics[width=1.0\linewidth]{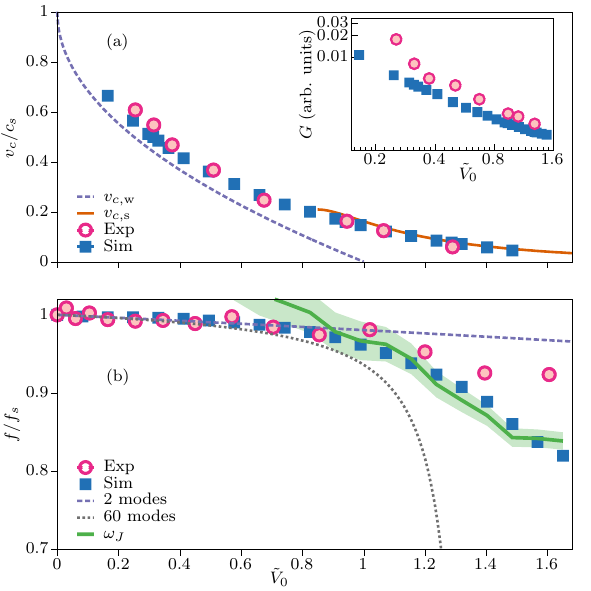}
\caption{Weak-link versus tunnel regimes. 
(a) Measured critical velocity (circles) and simulation results (squares), both normalized by the sound velocity $c_s$, 
are compared with theoretical predictions. 
The weak-link prediction $v_{c, \mw}$ (dashed line) is derived  for thick rectangular barriers \cite{Watanabe2009, Leszczyszyn2009}, 
while the ideal junction prediction $v_{c, \ms}$ (continuous line) is based on the tunneling amplitude $t_0( \tilde{V}_0, w)$. 
Inset shows the corresponding conductance $G$ plotted on a log-log scale. 
(b) Josephson oscillation frequency $f$, normalized by the sound frequency $f_s$, as a function of barrier height $\tilde{V}_0$. 
Hydrodynamic predictions using two modes (dashed line) and $60$ modes (dotted line) are shown.
The solid line represents the Josephson frequency prediction $\omega_J= \sqrt{I_c/(\hbar C)}$, 
using numerically determined critical current $I_c$ and capacitance $C$; see text for details. 
The shaded region indicates the standard error. 
Fitting errors of $v_c$ and $f$ are smaller than the symbol size. 
}
\label{Fig:vc}
\end{figure}

Experiments are performed with a BEC of about $180,000$  $^{87}$Rb atoms confined in a tube-like cylindrical geometry with harmonic
transverse confinement. 
For experimental details, see Ref. \cite{Bernhart2025}. 
The trapping frequencies are $(\omega_x, \, \omega_y, \, \omega_z) = 2\pi \times (1.6,\, 252, \,250)\, \mHz$. 
To achieve an approximately uniform density profile along the $x$ direction, 
the cloud is confined to $x=\pm \SI{37.5}{\um}$ using two optical end caps. 
A movable optical barrier and the two confining end-caps are realized with tightly focused laser beams. 
The condensate is split into two reservoirs (left and right) connected via a potential barrier. 
The barrier width is $w \approx \SI{1.1}{\um}$, and the barrier height $V_0$ is tunable, 
enabling exploration of both the weak-link and tunneling regimes. 
To induce a dc atom current, the barrier is moved at a constant velocity $v$, 
implemented as a time-dependent position $x(t) = v t$ \cite{Bernhart2025, DelPace2025}, 
with a fixed drive time of $t=33\, \mms$. 
Using absorption imaging, we measure the atom number imbalance $z= (N_R - N_L)/N$, 
where $N$ is the total atom number, and $N_L\, (N_R)$ is the atom number in the left (right) reservoir [Fig. \ref{Fig:system}(a)]. 
Since the final barrier position depends on the velocity $v$, we also determine a reference imbalance $z_\mathrm{ref}$ (measured without the barrier), and define the  imbalance difference $\Delta z = z - z_\mathrm{ref}$ as a function of $v$. 
This protocol, previously used for atomic Josephson junctions (AJJs) in Fermi gases \cite{Kwon2020, Pace2021}, allows us to extract the velocity-chemical potential relation, which is analogous to the current-voltage relation of superconducting Josephson junctions.
We numerically solve the Gross-Pitaevskii equation using imaginary time propagation to obtain 
the equation of state for the chemical potential: $\mu_0(N)=4.619\, \mHz \times \sqrt{N}$ at the trap center \cite{Bernhart2025}.
With this, and the measured $\Delta z$, we calculate the chemical potential difference
$\Delta \mu (\Delta z, N)= \mu(N\cdot(1+ \Delta z)) - \mu(N\cdot(1- \Delta z))$  \cite{Giovanazzi2000, SinghShapiro}. 
The resulting velocity-chemical potential characteristics are shown in Fig. \ref{Fig:system}(b).
For complementary measurements, we excite Josephson oscillations by introducing a small initial imbalance between the reservoirs.
The protocol involves ramping up a strong barrier slightly off-center, 
slowly translating it to the trap center, and then quickly ramping it down to the desired height. 
This excites oscillations in the imbalance $z(t)$ with initial amplitudes of about $2-6\%$, 
from which we extract the oscillation frequency $f$ as a function of $V_0$ [see Fig. \ref{Fig:vc}(b) and Appendix \ref{App:osc}].  
Since the system operates in the 3D regime, i.e., $\mu_0 > \hbar \omega_y,  \hbar \omega_z$, and the transverse confinement is harmonic, for $V_0$ near $\mu_0$, parts of the cloud are in the tunneling regime and parts are in the hydrodynamic regime. Increasing the barrier height shifts the junction more and more into the tunneling regime. To account for this, we introduce a  trap-averaged chemical potential $\mu= (2/3)\mu_0$, 
where the factor $2/3$ arises from spatial averaging over the transverse Thomas-Fermi profiles. 
We then define the normalized barrier height as $\tilde{V}_0 \equiv V_0/\mu$.

\begin{figure}[t]
\includegraphics[width=1.0\linewidth]{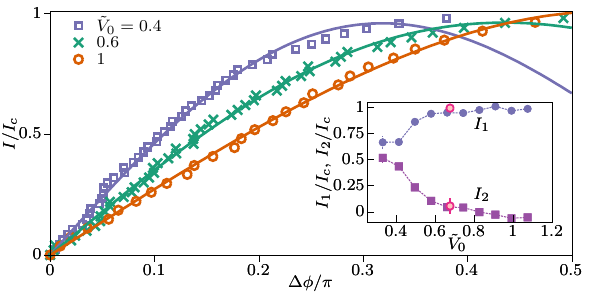}
\caption{Current-phase relation (CPR) at different barrier heights $ \tilde{V}_0$. 
The CPR is extracted from the same simulations as in Fig. \ref{Fig:vc}(a), 
and the phase difference is defined as $\Delta \phi = \phi(I)-\phi(0)$. 
Solid lines represent fits to the form  $I(\phi)= I_1 \sin(\phi) + I_2 \sin(2\phi)$, 
with $I_1$ and $I_2$ as free parameters.
Inset shows the extracted values of  $I_1$ and $I_2$ as a function of $ \tilde{V}_0$, 
where the experimental data  at $ \tilde{V}_0 = 0.67$ is from Ref. \cite{Bernhart2025}. 
}
\label{Fig:cpr}
\end{figure}

\begin{figure*}
\includegraphics[width=1.0\linewidth]{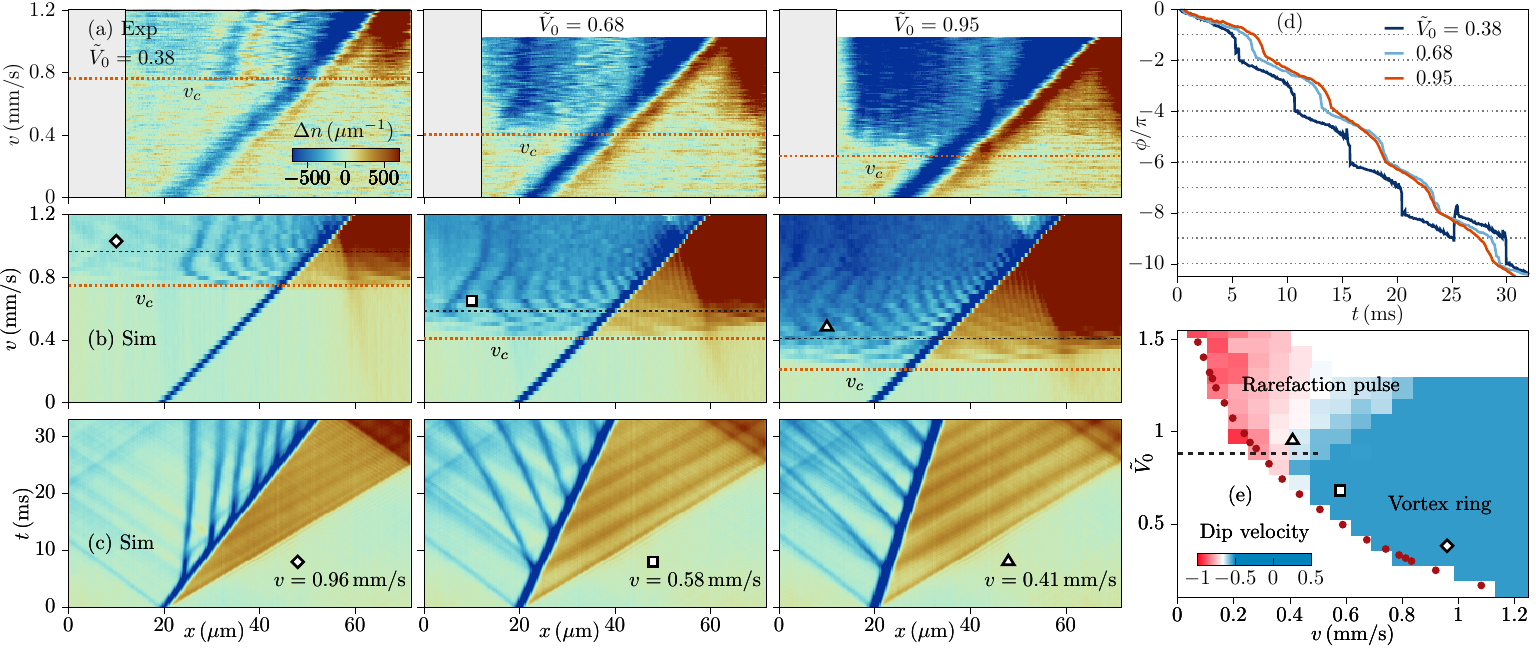}
\caption{Dynamics of density and phase. 
Time evolution of the density variation $\Delta n(x, t)= n(x, t) - n(x)$ in the experiment (a) and in the simulations (b, c), 
where $n(x)$ denotes the equilibrium line density without the barrier. 
(a, b) $\Delta  n(x, t=33\, \mms)$, taken at the end of the protocol, shown as a function of the barrier velocity $v$ for different barrier heights $\tilde{V}_0=0.38$, $0.68$, and $0.95$. 
The horizontal dashed line indicates the critical velocity $v_c$ marking the onset of the dissipative regime.  
The gray shaded region denotes the vacuum in the experiment. 
(c) $\Delta n(x, t)$ for selected velocities $v$, indicated by the horizontal dashed lines in the dissipative regime of (b).  
(d) Unbounded phase difference across the barrier, $\phi(t)$, exhibiting discrete  phase slips of value $\pi$ during the time evolution of a single trajectory. 
For $\tilde{V}_0=0.38$, the phase-slip dynamics becomes sufficiently rapid that finer numerical time sampling is required to resolve each individual slip; as a consequence, one slip is not resolved near $t \approx 25\, \mms$. 
(e) Extracted density dip velocity $v_e$ in the rest frame of the condensate, normalized by the sound velocity $c_s$. 
According to the Jones–Roberts theory of solitary waves, excitations with $|v_e|/c_s \lesssim 0.66$ correspond to vortex rings, 
whereas excitations at higher velocities correspond to rarefaction pulses. 
The horizontal dashed line indicates the crossover inferred from the vorticity analysis (Appendix \ref{App:cros}).
Dots indicate $v_c$, and symbols correspond to the results shown in panels (c).
}
\label{Fig:dynamics}
\end{figure*}

\begin{figure}
\includegraphics[width=0.95\linewidth]{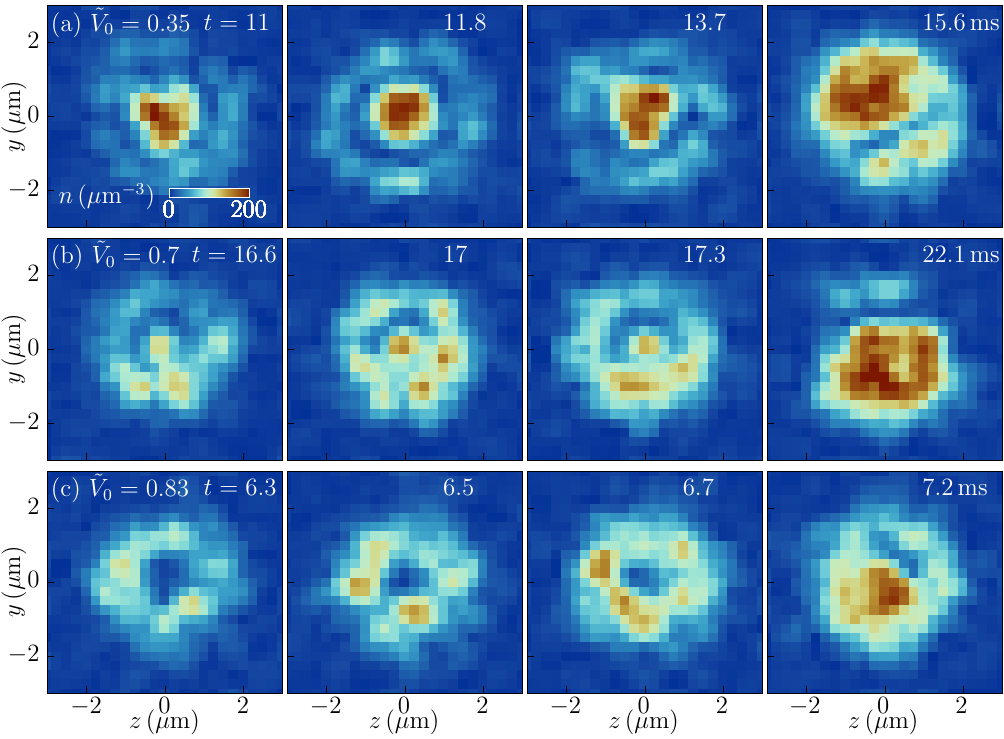}
\caption{Nucleation of vortex rings. 
(a-b) Transverse density slices $n(y,z)$ at different times $t$, showing nucleation and decay of vortex rings after a phase slip.  
The location $x(t)$ of the slices follows the propagation of the low-velocity density pulse identified in the column density evolution. 
(a) $\tilde{V}_0=0.35, \, v=0.9\, \mathrm{mm/s}$. 
(b) $\tilde{V}_0=0.7, \, v=0.5\, \mathrm{mm/s}$. 
(c) $\tilde{V}_0=0.83, \, v=0.5\, \mathrm{mm/s}$. 
}
\label{Fig:ring}
\end{figure}
\section{Weak-link versus tunnel-junction regimes}
In Fig. \ref{Fig:system}(b), the measured  $v-\Delta \mu$ relations for different $\tilde{V}_0$ 
resemble the characteristic behavior of superconducting Josephson junctions. 
At low velocities, a dissipationless dc Josephson effect is observed, 
while above a critical velocity $v_c$, a finite chemical-potential difference $\Delta \mu$ emerges across the junction, 
signaling  a transition to the resistive ac Josephson regime.  
The critical velocity $v_c$ decreases with increasing barrier height $\tilde{V}_0$. 
To quantify this behavior, we fit the data using the resistively shunted junction (RSJ) circuit model, 
which predicts $\Delta \mu = G^{-1}\sqrt{v^2 - v_c^2}$, where $G$ is the conductance and $v_c$ the critical velocity, 
both treated as fit parameters \cite{SinghShapiro}. 
This model captures the dynamics of the junction and enables a reliable determination of $v_c$ and $G$.  
To corroborate the measurements and gain microscopic insight, we simulate the system dynamics using a classical-field method 
within the truncated Wigner approximation \cite{Blakie2008, Polkovnikov2010, Singh2016, Singh2020sound}, see Appendix \ref{App:sim} for details. 
Following the same protocol as in the experiment,  we determine the  $v-\Delta \mu$ relations from simulations, shown in Fig. \ref{Fig:system}(c). 
The simulation results reproduce the dc-ac transition and closely follow the experimental trends.

From the data in Fig. \ref{Fig:system}, we extract the critical velocity $v_c$ and the conductance $G$ both from experiment and from theoretical modeling. The results are presented in Fig. \ref{Fig:vc}(a).
 We find excellent agreement between the measured and simulated values of $v_c$, 
 which decrease systematically with increasing barrier height $\tilde{V}_0$.
The decrease is rapid for low to intermediate $\tilde{V}_0$ and slows at higher values.  
For thick rectangular barriers ($d/\xi \gg 1$), 
hydrodynamic theory  predicts  $v_{c,\mw}/c_s \approx 1- \sqrt{3V_0/(2\mu_0)}$ \cite{Watanabe2009, Leszczyszyn2009}, 
where $d \approx w$ is the barrier width, $\xi \approx  \SI{0.3}{\um}$ is the healing length, and $c_s \approx 1.68 \, \mathrm{mm/s}$ is the transversely averaged sound velocity at the trap center.  
This prediction captures the behavior up to  $\tilde{V}_0 \sim 0.7$.
For larger $\tilde{V}_0$, the slower decrease in $v_c$ signals entry into the tunneling regime, 
where $v_c$ follows the tunneling amplitude $t_0(V_0, d)$ of an ideal AJJ, with $v_{c, \ms}/c_s = t_0(V_0, d)$  \cite{Singh2020jj}.
This result agrees with both measured and simulated $v_c$ for all $\tilde{V}_0 \gtrsim 0.8$. 
The extracted conductance $G$ also decreases with increasing  $\tilde{V}_0$ [Fig.~\ref{Fig:vc}(a), inset], 
consistent with quantum point contact behavior $G \propto v_c^2$ (Appendix \ref{App:cond}).

Figure \ref{Fig:vc}(b) shows the Josephson oscillation frequency $f$ from both experiment and simulation. 
In the weak-link regime, $f$ remains close to the sound mode frequency $f_s=c_s/(2L_x)$, with $L_x=\SI{75}{\um}$ being the system length in $x$ direction,  indicating that a bulk-dominated dynamics occurs. 
Treating the Gaussian barrier as a perturbation, 
hydrodynamic theory \cite{Pethick_Smith_2008} yields the sound frequency for the lowest antisymmetric mode across the barrier 
(Appendix \ref{App:pred})
\begin{align}
\frac{f^2}{f_s^2} \simeq 1 - \frac{\sqrt{2\pi} V_0 w}{\mu L_x}.
\end{align}
This low-$V_0$ expression explains the slow decrease of $f$. 
At larger $V_0$,  we solve the hydrodynamic equations numerically by including up to $60$ modes. This multimode calculation  
provides a further renormalization of the mode frequency and 
reveals a sharper reduction near $\tilde{V}_0 \sim 1$. 
In the tunneling regime, the Josephson frequency is expected to follow $\omega_J= \sqrt{I_c/(\hbar C)} $,
where $I_c$ is the critical current and $C$ is the capacitance \cite{tinkham2004introduction}. 
By extracting $I_c$ and $C$ from the time evolution of imbalance and junction phase (Appendix \ref{App:sim}), 
we find good agreement with both simulation and experiment for $\tilde{V}_0 \gtrsim 0.9$,
except near $\tilde{V}_0 = 1.6$. 
In this region, the measurement of the oscillation frequency becomes less reliable
because the amplitude of Josephson oscillations is small, leading to a low signal-to-noise ratio (Fig. \ref{AppFig:ztime}).

  To analyze the onset of the tunneling regime, we examine the current-phase relation (CPR) from simulations. 
For each barrier velocity $v$, we determine the phase $\phi(t=33\, \mms)$ across the barrier from the wave function $\psi(\br, t)$ and average it over the initial ensemble, where $v$ is equivalent to the current, i.e., $I/I_c = v/v_c$. 
In Fig. \ref{Fig:cpr}, the CPR shows a transition from second-harmonic to first-harmonic behavior. 
Around $\tilde{V}_0 \sim 0.8$, the first harmonic becomes dominant while the second harmonic is strongly suppressed, consistent with the onset of the tunneling regime.

\section{Vortex-ring to rarefaction-pulse dissipation}
To elucidate the microscopic origin of resistive dynamics in the AC Josephson regime, we analyze  the time evolution of the density variation $\Delta n(\br, t)$ in both experiment and simulations.
The density maps in Figs. \ref{Fig:dynamics}(a, b) reveal two distinct transport regimes as the barrier velocity $v$ is increased.
Below the critical velocity, the density remains essentially unchanged across the junction, 
indicating dissipationless transport. 
Above this threshold, a dissipative regime emerges, characterized by a density depletion on one side of the barrier 
and a corresponding accumulation on the other.
Within the dissipative regime, structured low-density regions develop near the junction. 
While these features are not clearly visible in the experiment due to finite imaging resolution, they are clearly resolved in the simulations, where localized excitations are generated during the barrier motion and propagate into the bulk [Fig. \ref{Fig:dynamics}(c)].
Their emission is directly correlated with the phase dynamics at the barrier: as the unbounded phase difference $\phi(t)$ decreases, discrete phase slips of value $\pi$ occur [Fig. \ref{Fig:dynamics}(d)], marking the onset of resistive dynamics. 
For fixed barrier height and velocity, $\phi(t)$ decreases deterministically at a rate $\dot{\phi} \simeq -\Delta\mu/\hbar$, leading to phase slips that occur at well-defined times when the accumulated phase reaches a critical value, rather than arising from stochastic fluctuations.
Each phase slip produces both phononic and solitonic excitations, the latter manifesting as density dips propagating with velocity smaller than phonon velocity, see Fig. \ref{Fig:dynamics}(c). 
These solitonic excitations emerge as vortex rings (VRs) for low and intermediate barrier heights, and as rarefaction pulses (RPs) for strong barrier heights, which we have identified from 3D density isosurfaces and phase-circulation analysis, yielding the VR-RP crossover at $\tilde{V}_0 \sim 0.9$ (Appendix \ref{App:cros}).
By tracking the propagation velocity $v_e$ of the density dips, 
we construct a dynamical phase diagram as a function of $v$ and $\tilde{V}_0$ in Fig. \ref{Fig:dynamics}(e). 
Based on the Jones-Roberts theory of solitary waves, excitations with $|v_e|/c_s \lesssim 0.66$ correspond to VRs, 
while excitations at higher velocities are expected to be  RPs \cite{Natalia2004, Myrann2025}. 
This criterion is in quantitative agreement with the VR--RP crossover inferred from the barrier-height dependence [Fig. \ref{Fig:dynamics}(e)].  In Fig. \ref{Fig:ring}, transverse density slices directly reveal the nucleation of VRs for low and intermediate barrier heights, which rapidly decay into vortex lines (Appendix \ref{App:cros}). 
For larger barrier heights, VR formation is suppressed and the excitations instead take the form of RPs, reflecting both the reduced effective junction cross section and the inability of fast excitations to sustain quantized circulation. 
This crossover is accompanied by a broadening of the phase slips.

Taken together, these observations establish a unified microscopic picture in which dissipation in the AC Josephson regime arises from phase-slip-induced emission of solitonic excitations, with a crossover from VR-dominated dissipation to RP emission as the barrier height is increased. 
Solitonic structures have been observed in BECs created via phase imprinting \cite{Ku2014} or rapid quenches \cite{Donadello2014}.  
In thin AJJs realized with $^{6}$Li superfluids, dissipation was also linked to phase slips involving vortex rings \cite{Burchianti2018, Xhani2020, Xhani2020NJP},  
consistent with our scenario of weak-link-induced dynamics.

\section{Conclusions and outlook}
We have studied the driven Josephson dynamics in both the weak-link and tunneling regimes, as well as the crossover between them, in a BEC of $^{87}$Rb atoms separated by an optical barrier.
Our results provide direct insight into the microscopic mechanisms governing  coherent transport and dissipation in quantum circuits.  

In our system, both the reservoirs and the junction barrier are  three dimensional (see Fig.\ref{Fig:system}), giving rise to characteristic features of the Josephson dynamics.  
We quantify how the weak-link and tunneling regimes are characterized by markedly different critical currents, Josephson frequencies and current-phase relations (see Figs. \ref{Fig:vc} and \ref{Fig:cpr}).
In the weak-link regime, superfluid flow above the critical velocity generates vortex-ring excitations that mediate phase slips and dissipation--see Figs. \ref{Fig:dynamics} and \ref{Fig:ring}. 
Increasing the barrier height effectively shrinks the junction, suppressing vortex-ring formation, and driving the system into a tunneling-dominated regime in which dissipation is instead associated with rarefaction-pulse excitations. Measurements of the critical currents and Josephson oscillations, supported by simulations, reveal a smooth crossover between these regimes in agreement with theoretical models.

Given the level of control demonstrated here, our system provides a versatile platform for exploring the dynamics of 3D vortices and  turbulence in quantum fluids \cite{Singh2025VR}. In particular, questions related to energy cascades, vortex reconnections, dissipation mechanisms, and the quantum-classical crossover  \cite{barenghi2023quantum} can be addressed under well-defined and tunable physical conditions.

\section*{Acknowledgments}  
L. M. acknowledges funding by the Deutsche Forschungsgemeinschaft (DFG) in the excellence cluster  `Advanced Imaging of Matter’ - EXC 2056 - project ID 390715994 and by ERDF of the European
Union and `Fonds of the Hamburg Ministry of Science,
Research, Equalities and Districts (BWFGB)’. H.O., E.B., and M.R. acknowledge financial support by the DFG within the SFB OSCAR (project number 277625399).


\appendix
\renewcommand{\thefigure}{A\arabic{figure}}
\renewcommand{\theequation}{A\arabic{equation}}
\setcounter{figure}{0}
\setcounter{equation}{0}

\section{Josephson oscillations}\label{App:osc} 
To probe Josephson oscillations, we prepare the system with a controlled initial atom number imbalance $z(0)$ between the reservoirs. 
This is achieved by first ramping up a strong optical barrier slightly off-center in the trap, thereby creating an atom number imbalance. 
The barrier is then slowly translated to the trap center and rapidly lowered to the target barrier height $\tilde{V}_0$. 
This sudden change triggers oscillation of the imbalance $z(t)$, which is recorded by taking absorption images at various hold times after the excitation.
Figure \ref{AppFig:ztime}(a) shows measured $z(t)$ for two different values of $\tilde{V}_0$. 
Fitting these oscillations with a damped sinusoidal function allows us to determine the oscillation frequency $f$ as a function of $\tilde{V}_0$. 
The initial amplitude $z(0)$ falls in the range between $2-6\%$. 
To compare these measurements we perform numerical simulations using the experimental parameters and the excitation protocol, 
as described in the next section.
The simulation results of $z(t)$ are presented in Fig. \ref{AppFig:ztime}(b), where the initial imbalance $z(0)$ lies in the range between $1.5-4\%$.

\section{Simulation method}\label{App:sim} 
To benchmark experimental results and gain microscopic insights, we simulate the dynamics of the system using a classical-field method 
within the truncated Wigner approximation \cite{Blakie2008, Polkovnikov2010, Singh2016, Singh2020sound}.  
The system is described by the Hamiltonian
\begin{align} \label{eq:hamil}
\hat{H} &= \int \mathrm{d}{\bf r} \Big[  \frac{\hbar^2}{2m}  \nabla \hat{\psi}^\dagger({\bf r}) \cdot \nabla \hat{\psi}({\bf r})  + V_\mh({\bf r}) \hat{\psi}^\dagger({\bf r})\hat{\psi}({\bf r})  \, \nonumber  \\
&    \quad + \frac{g}{2} \hat{\psi}^\dagger({\bf r})\hat{\psi}^\dagger({\bf r})\hat{\psi}({\bf r})\hat{\psi}({\bf r})\Big],
\end{align}
$\hat{\psi}$ ($\hat{\psi}^\dagger$) is the bosonic annihilation (creation) operator. 
The 3D interaction parameter is given by $g=4\pi a_s \hbar^2/m$, 
where $a_s=5.3\, \mathrm{nm}$ is the $s$-wave scattering length and $m$ is the atomic mass. 
The external potential $V_\mh({\bf r})$ represents the harmonic trap with frequencies same as the experiment. 
In the classical-field approximation we replace the operators $\hat{\psi}$ in Eq. \ref{eq:hamil} and in the equations of motion by complex numbers $\psi$.  We map real space on a lattice system of  $320 \times 25 \times 25$ sites with the discretization length of $\SI{0.25}{\um}$. 
We sample the initial states $\psi (\br,t=0)$ in a grand canonical ensemble with chemical potential $\mu$ and temperature $T$ via a classical Metropolis algorithm. 
We use $T= 23\, \mathrm{nK}$ and adjust $\mu$ such that the total atom number $N$ matches the experimental value. 
The initial ensemble of $\psi (\br,t=0)$ includes thermal fluctuations, thus closely describing the finite-temperature experiment. 
Finite-temperature effects are therefore essential for a quantitative comparison with the experiment, even though the qualitative features are already present at zero temperature.
Each initial state is subsequently evolved using the equations of motion
\begin{equation}\label{eq:eom}
 i \hbar \dot{\psi}(\br, t) = \Bigl(  - \frac{\hbar^2}{2m} \nabla^2 +  V_\mh(\br, t) +  V_\mb(\br, t) + g|\psi|^2 \Bigr) \psi(\br, t),
\end{equation}
which include the barrier potential given by $V_\mb({\bf r},t)  = V_0 (t) \exp \bigl[- 2 \bigl( x-x(t) \bigr) ^2/w^2 \bigr]$. $V_0(t)$, $w$  and $x(t)$ are the barrier's  strength, width and location.  
Following experimental protocol, we slowly turn-on the barrier height 
and then move the barrier at velocity $v$  via $x(t)= vt$ for $t=33\, \mms$.  
From the time evolution $\psi (\br,t)$, we calculate the density distribution $n(\br,t)=|\psi (\br,t)|^2$ and the phase distribution  $\phi(\br,t)$. 
At the end of the protocol, we compute the imbalance $\Delta z$ using atom numbers of left and right reservoirs. 
We obtain the chemical-potential difference $\Delta \mu (\Delta z, N)$ as a function of velocity $v$, see Fig. \ref{Fig:system}(c). 

\begin{figure}[t]
\includegraphics[width=1.0\linewidth]{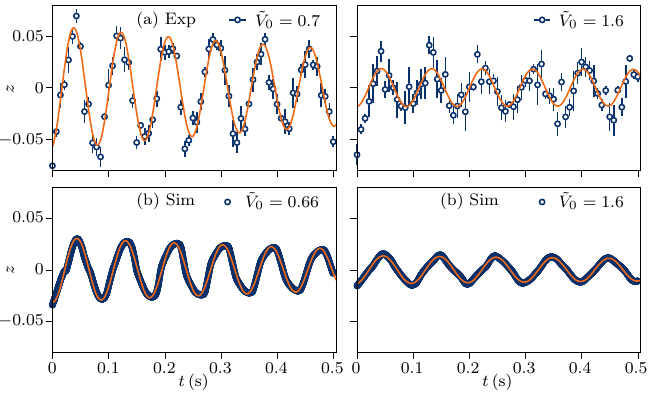}
\caption{Josephson oscillations. (a) Measured time evolution of the imbalance $z(t)$ for barrier heights $\tilde{V}_0= 0.7$ and $1.6$. 
The solid lines are the fits to a damped sinusoidal function to determine the oscillation frequency, the amplitude and the damping rate. 
(b) Simulation of the time evolution of the imbalance for parameters close to the experiments.
}
\label{AppFig:ztime}
\end{figure}

\begin{figure}[t]
\includegraphics[width=1.0\linewidth]{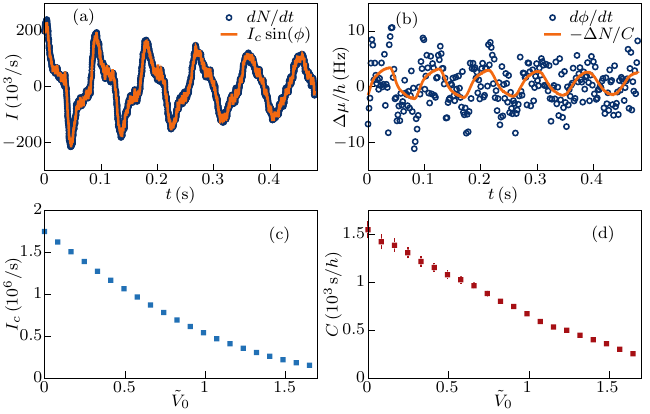}
\caption{
(a, b) Atom number difference $\Delta N(t)$ and the phase difference $\phi(t)$ across the barrier of Josephson oscillations and their numerical derivatives $d N/dt$ and $d \phi/dt$ allow us to extract the critical current $I_c$ and the capacitance $C$ following the relations $d N/dt = I_c \sin \bigl( \phi(t) \bigr)$ and $d \phi/dt = - \Delta N /C$. For $\tilde{V}_0=1$, this yield $I_c = 537 \times 10^3/\mathrm{s}$ and $C= (671 \pm 25)\, \mathrm{s}/h$.  
(c, d) $I_c$ and $C$ as a function of $\tilde{V}_0$. 
}
\label{AppFig:Itime}
\end{figure}

To excite Josephson oscillations, we create an initial atom number imbalance $z(0)$ following the experimental procedure described above. 
The time evolution of the resulting imbalance  $z(t)$, determined using the atom numbers $N_L$ and $N_R$, 
exhibits Josephson oscillations, as shown in Fig. \ref{AppFig:ztime}(b). 
These oscillations are also present in the junction phase that we determine in the vicinity of the barrier using $\phi  = \phi(x+2l) - \phi(x-2l)$, 
where $x=L_x/2$ is the barrier location. 
For this, we average the field $\psi(\br, t)$ over $yz$ directions to take into account fluctuations due to inhomogeneity. 
We use $\Delta  N (t)= N_R - N_L$ and $\phi(t)$ to determine the critical current $I_c$ and the capacitance $C$ of the junction \cite{Bernhart2025}. 
We apply a Gaussian filter and calculate the current $I(t)= dN/dt$ by numerical differentiation of $\Delta  N (t)$, 
which we fit with the ideal junction relation $I(t) = I_c \sin \bigl(  \phi(t)  \bigr)$ to determine the only free parameter $I_c$, see Fig. \ref{AppFig:Itime}(a).
 
 With $\Delta  N (t)$ and $\phi(t)$, we can also estimate $C$. Based on the Josephson relation $\hbar \dot{\phi}(t) = - \Delta \mu$,
 numerical derivative of $\phi(t)$ gives $\Delta \mu(t)  = d \phi(t)/dt $, which we fit to $-\Delta N(t)/C$ using $C$ as the only fitting parameter, see Fig. \ref{AppFig:Itime}(b). This way, we determine $I_c$ and $C$ as a function of barrier height $\tilde{V}_0$ [Fig. \ref{AppFig:Itime}(c,d)].

\begin{figure}
\includegraphics[width=1.0\linewidth]{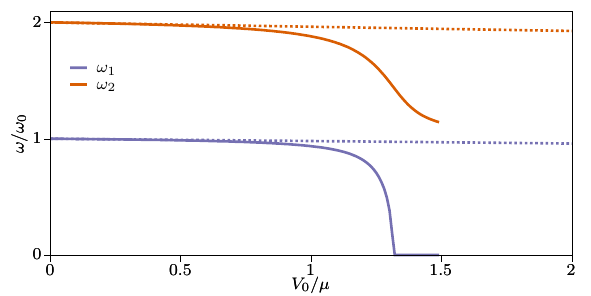}
\caption{Frequencies of the modes of the two-mode model (dashed lines) given by Eq.  \ref{eq:freq}, as a function of barrier height $V_0/\mu$. 
Numerical results for the lowest two modes (solid lines) using $60$ modes and the full expression for $M^V$ in Eq. \ref{eq:barFull}. 
We use the healing length $\xi \approx  \SI{0.3}{\um}$ and the barrier width $\sigma = w/2=\SI{0.55}{\um}$. 
}
\label{AppFig:hd}
\end{figure}

\begin{figure}
\includegraphics[width=1.0\linewidth]{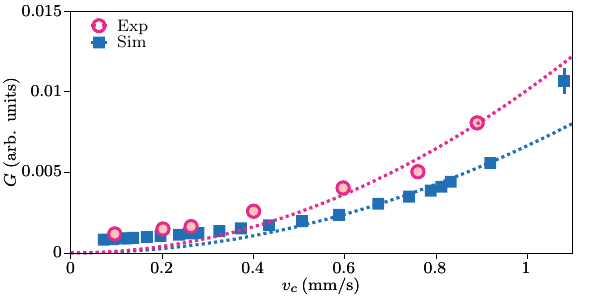}
\caption{Measured conductance $G$ (circles) and simulation results (squares) shown as a function of the critical velocity $v_c$.
The $y$ error bars correspond to the fitting errors. 
The results are fitted with the function $G=\alpha v_c^2$ (dashed lines), with $\alpha$ being the free parameter.
}
\label{Fig:Gvc}
\end{figure}

\begin{figure*}
\includegraphics[width=1.0\linewidth]{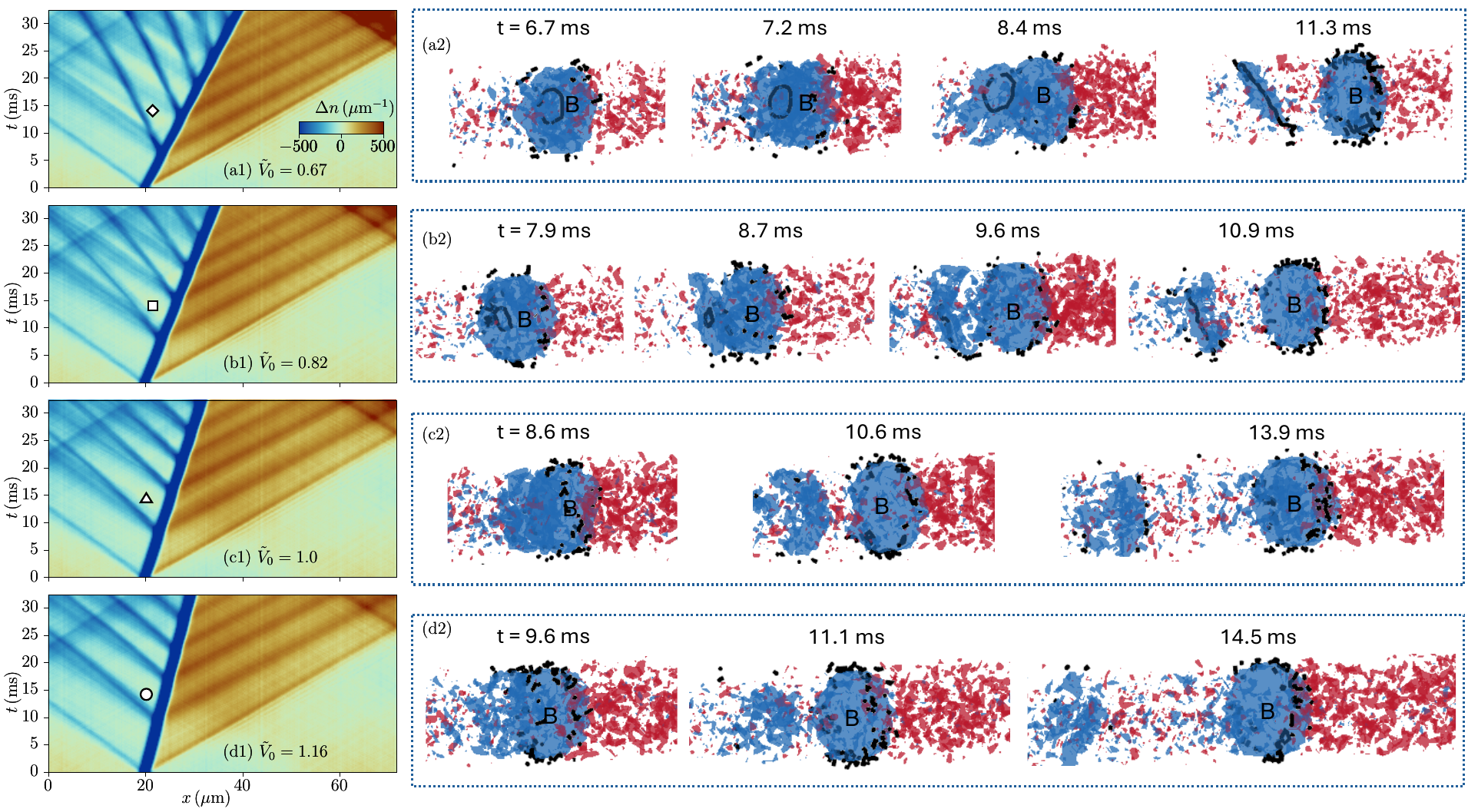}
\caption{Dynamics of density excitations in the dissipative AC Josephson regime.
(a1--d1) Time evolution of the density variation $\Delta n(x,t)=n(x,t)-n(x)$ obtained from simulations for different barrier heights in the range $\tilde{V}_0=0.67$--$1.16$,
where $n(x)$ denotes the equilibrium line density in the absence of the barrier.
(a2--d2) Time-resolved snapshots of 3D density-difference isosurfaces together with the extracted phase vorticity, rendered as tubular filaments, corresponding to the excitation dynamics indicated by the symbols in panels (a1--d1).
Closed vorticity loops correspond to vortex rings, while open filaments indicate vortex lines.
The label B marks the density depletion at the barrier. 
Additional isolated vortices appear in the vicinity of the barrier due to enhanced phase fluctuations and local density suppression at large $\tilde{V}_0$.}
\label{Fig:VR-RP}
\end{figure*}

\section{Hydrodynamic prediction for Josephson oscillation frequency}\label{App:pred} 
We use hydrodynamic theory to determine the spectrum of collective modes in a box potential with a Gaussian barrier. 
Within the Gross-Pitaevskii equation in terms of the velocity $v(\br, t)$ and density field $n(\br, t)$, we obtain a set of coupled equations \cite{Pethick_Smith_2008}. 
For the density dynamics, adjusted to one-dimensional motion, linearlized hydrodynamic equation yields 
\begin{equation}
\omega^2 n = - \frac{\mu}{m} \partial_{xx} n + \frac{\hbar^2}{4m^2} \partial_{xxxx} n +   \frac{(\partial_x V) (\partial_x n) + V \partial_{xx} n}{m}.
\end{equation}
On the right hand side, the $\mu$-term describes sound-wave propagation, 
the second term comes from quantum pressure, 
and the third term includes the effect of barrier. 
We use the following functions as a basis on the interval $x \in [-L, L]$:
\begin{align}
n_p(x) = \frac{1}{\sqrt{L}} \sin(\pi p x /L),
\end{align}
with $p=1, 2, ...$. The barrier potential is  $V(x) = V_0  \exp \bigl( - x^2 /(2\sigma^2) \bigr)$. 
In this basis, the terms of the hydrodynamic  equation are
\begin{align}
\sum_{p_1} M^0_{p_2, p_1} n_{p_1} = \sum_{p_1} \bigl(  M^{\mu}_{p_2, p_1}  + M^4_{p_2, p_1}  + M^V_{p_2, p_1}   \bigr) n_{p_1}, 
\end{align}
with the expressions: 
\begin{align}
M^0_{p_2, p_1}  &=  \omega^2 \delta_{p_1, p_2} \\
M^{\mu}_{p_2, p_1}  &= \omega_0^2 p_1^2 \delta_{p_1, p_2}  \\
M^4_{p_2, p_1}   &=   \omega_0^2   \frac{\pi^2 \xi^2}{2L^2}  p_1^4 \delta_{p_1, p_2}  \\ 
M^V_{p_2, p_1} &= - \frac{V_0}{\mu} \frac{\sqrt{2\pi} \sigma}{L} \omega_0^2   p_1 p_2   \cosh \Bigl(  \frac{p_1 p_2 \pi^2 \sigma^2}{L^2}    \Bigr) \\
                          & \quad \exp \Bigl( - \frac{ (p_1^2 + p_2^2) \pi^2 \sigma^2 }{2L^2}   \Bigr)   \label{eq:barFull} \\
                          & \approx   - \frac{V_0}{\mu}  \frac{\sqrt{2\pi} \sigma}{L} \omega_0^2   p_1 p_2   \label{eq:bar}
\end{align}
with $\omega_0^2 = \mu \pi^2/(m L^2)$. 
Taking only $p= 1, 2$ into account, using the approximation in Eq. \ref{eq:bar}, and at fixed chemical potential:
\begin{align}
\frac{\omega_{1/2}^2}{\omega_0^2} &= \frac{5-5x \mp \sqrt{9-18x+25x^2}}{2} ,  \label{eq:freq}
\end{align}
where $x= V_0 \sqrt{2\pi} \sigma/(\mu L)$.  
This results in the ground state frequency for the lowest mode 
\begin{align}
\frac{\omega_1^2}{\omega_0^2} &= \frac{5-5x - \sqrt{9-18x+25x^2}}{2}   \label{eq:freq_gs}\\
& \approx 1 - x - \frac{4 x^2}{3}.
\end{align}
In Fig. \ref{AppFig:hd}, we show the analytical result for the two-mode approximation. 
By including the lowest $60$ modes, $p=1,2, 3, ...,60$, and using the full expression for $M^V$, shown in Eq. \ref{eq:barFull}, results in a further renormalization of the mode frequencies.

\section{Conductance--critical-velocity dependence}\label{App:cond}
In this section, we examine the dependence of the junction conductance $G$ on the critical velocity $v_c$, with both quantities extracted from the velocity--chemical-potential characteristics of the junction.
Figure~\ref{Fig:Gvc} shows the experimentally measured and numerically simulated conductance as a function of $v_c$.
Both results display a clear nonlinear increase of $G$ with increasing $v_c$. Within the framework of a quantum point contact, the conductance is expected to scale quadratically with the critical velocity, $G \propto v_c^2$ \cite{Uchino2020}. 
To quantify this behavior, we fit the results using the functional form $G= \alpha v_c^2$, 
where $\alpha$ is a single fitting parameter.
This quadratic dependence captures the overall trend of both the experimental and simulation results. The extracted values of the prefactor are $\alpha = 0.007$ for the simulations and $\alpha = 0.01$ for the experiment, indicating quantitative agreement between the two within experimental and fitting uncertainties.

%
%

\section{Vortex-ring to rarefaction-pulse crossover}\label{App:cros}
In this section, we expand on the dissipation mechanism that emerges in the AC Josephson regime as the barrier height $\tilde{V}_0$ is increased. As demonstrated in the main text, the dissipative regime is characterized by phase-slip dynamics, which leads to the generation of both phononic and solitonic excitations. The latter manifest as localized density dips propagating at velocities below the phonon velocity. 
To characterize these solitonic excitations, we analyze the time evolution of three-dimensional (3D) density isosurfaces together with the associated  phase vorticity. 
The phase vorticity is computed following the method described in Ref. \cite{Singh2025VR}. 
Vortex filaments are identified from the complex order parameter $\psi(\mathbf{r},t)$ using a phase-circulation (plaquette) criterion.
For each elementary grid face, we evaluate the wrapped phase circulation
$\Gamma_{\Box} = \sum_{j=1}^{4} \Delta \phi_j$, where the phase increments
$\Delta \phi_j = \phi_{j+1} - \phi_j$ are taken modulo $2\pi$ and restricted to the interval $(-\pi,\pi]$.
Grid faces satisfying $|\Gamma_{\Box}| > \pi/2$ are tagged as being pierced
by a vortex filament.

The vortex-core position is subsequently refined to sub-grid accuracy by
locating the intersection of the nodal surfaces
$\Re\,\psi = 0$ and $\Im\,\psi = 0$ using bilinear interpolation within each
flagged cell.
Intersection points lying within a prescribed spatial tolerance are paired
to form short filament segments, which are then merged and connected into
extended filaments by traversing an adjacency graph.
Closed filaments are identified as vortex rings (VRs).
Filaments that do not form closed loops are classified as vortex lines (VLs).
To suppress boundary artefacts arising from low-density regions, the analysis
is restricted to the interior of the Thomas--Fermi radii in the transverse
($y$--$z$) directions.

In Figs. \ref{Fig:VR-RP}(a2-d2), we present time-resolved snapshots of the 3D density difference isosurfaces together with the extracted vortex filaments. For barrier heights $\tilde{V}_0 \leq 0.82$, we find the emission of a VR at the barrier, which then propagates into the bulk and decays into a vortex line while emitting a phononic pulse, see Figs. \ref{Fig:VR-RP}(a2, b2). 
The VR lifetime is typically of order $1 \, \mathrm{ms}$ and depends on both the system temperature and inhomogeneity. 
At lower temperatures, the lifetime increases to several milliseconds, reaching values of order $5 \, \mathrm{ms}$ \cite{Singh2025VR}. 
In contrast, for barrier heights $\tilde{V}_0 \geq 1$,  the dominant excitation is a rarefaction pulse (RP) without any associated quantized circulation. These pulses propagate away from the barrier at higher velocities and disperse into phononic pulses, as shown in Figs. \ref{Fig:VR-RP}(c2, d2). By examining the intermediate regime between $\tilde{V}_0 = 0.82-1$, we identify a crossover barrier height $\tilde{V}_0 \sim 0.9$, marking the transition from VR emission to RP generation.

\bibliography{References}

\end{document}